\RequirePackage{fix-cm}
\documentclass{svjour3}                     
\smartqed  

\usepackage{color}
\usepackage{graphicx}
\usepackage{amsmath}
\usepackage{amsfonts}
\usepackage{amssymb }
\usepackage{float}
\usepackage{multicol}
\usepackage{array}
\usepackage{tabularx}

\usepackage{latexsym}

\makeatletter

\begin{document}
\title{Hidden transient chaotic attractors of Rabinovich-Fabrikant system}

\subtitle{}

\titlerunning{}

\author{Marius-F. Danca}

\institute{Marius-F. Danca \at
Department of Mathematics and Computer Science\\
Avram Iancu University, 400380 Cluj-Napoca, Romania \\
and \\
Romanian Institute for Science and Technology\\
400487 Cluj-Napoca, Romania \\
              \email{danca@rist.ro}}

\date{Received: date / Accepted: date}

\maketitle

\begin{abstract}

In \cite{dan}, it is shown that the Rabinovich-Fabrikant (RF) system admits self-excited and hidden chaotic attractors. In this paper, we further show that the RF system also admits a pair of symmetric transient hidden chaotic attractors. We reveal more extremely rich dynamics of this system, such as a new kind of ``virtual saddles''.

\end{abstract}

\section{Introduction}

Nowadays, the notion of self-excited and hidden attractor introduced by Leonov and Kuznetsov \cite{nick2,nick4,nick5}, has become a common subject (see e.g. \cite{kap,spry,xxx1,xxx2,xxx3,xxx4,x3x,xxx6,unu5,unu51,unu52,unu53}). The main characteristic of hidden attractors is that their basins of attraction do not intersect with arbitrary small neighborhood of any equilibrium point, while a basin of attraction of a self-excited attractor is associated with some unstable equilibrium. In this context, stationary points are less important for finding hidden attractors than for finding self-excited attractors. Self-excited attractors can be localized (excited) by standard computational procedures, by starting from a point in some neighborhood of an unstable equilibrium, while for localization of hidden attractors it is necessary to develop special numerical procedures. Some well-known classical chaotic and regular attractors (such as Lorenz\footnote{The possible existence in the Lorenz of a hidden chaotic attractor in the Lorenz system is an open problem \cite{nick5}.}, Chen, R\"{o}sler, van der Pol, some Sprott systems, etc.) are self-excited attractors, which can be localized numerically with standard computational procedures. However, as for the example considered in this paper, the system has both self-excited and hidden attractors. Hidden attractors are important in engineering applications because they may allow unexpected and potentially disastrous responses to perturbations in a structure like a bridge or aircraft wing.
Hidden attractors can be found in systems with no equilibria \cite{xxx0,unu}, or with only one stable equilibrium \cite{yyy1}, which is a special case of multistable systems or coexistence of attractors \cite{xx0}.
Uncovering all co-existing attractors and their underlying basins represents one of the major difficulties in locating hidden attractors.

On the other hand, transient chaos, ubiquitous in chaotic systems, is due to nonattracting chaotic saddles in phase space \cite{tr1,tr2,tr3,tr4,tr5,tr6,ttt,ttr1,ttr2,ttrx,treix,doix,patrux,sasex,asta}.
The transient chaos represents a phenomenon which appears when a nonlinear system behaves chaotically during some (short or long, but finite) transient time\footnote{Generally, the lifetime of the transient could be extremely long (superpersistent) (see \cite{tr5} where it is shown that the average lifetime of chaotic transients for two-dimensional maps could last tens of thousands of iterations and also scales with the system parameter variation).}, after which falls into a periodic or chaotic attractor. Initially, the behavior is aperiodic and the system is sensitive to initial conditions (i.e. ``chaos''), after which it settles down on a periodic orbit (or fixed point), or some chaotic attractor. Transient chaos is a common phenomenon of many engineering, physical and biological systems. Dynamics on systems with chaotic transients can be unpredictable even finally the system falls into a very simple motion. Chaotic transients can be found in systems with multiple (competing) attractors (as in the Rabinovich-Fabrikant (RF) system considered in this paper) and the existence of chaotic transients may have important implications for experiments on chaotic systems. For example, such transient phenomena were observed in hydrodynamics  \cite{doix}, radio circuits \cite{treix}, neural networks \cite{patrux}, Lorenz system \cite{sasex}, R\"{o}ssler system \cite{cincix}, maps \cite{asta}, experiments \cite{treix} and so on. Note that transient chaos can be quite disastrous and therefore unwanted, and it can be the cause of catastrophic developments in a dynamic system, for example in situations of voltage collapse or species extinction \cite{ttrx}. Therefore, control or anticontrol (in the sense of maintaining) of transient chaos can be desirable in some cases.

To note that in a recent paper \cite{mot} it is shown that a new phenomenon of doubly transient chaos, which is fundamentally different from the hyperbolic and nonhyperbolic transient chaos reported in the literature, appears in many other systems (chemical reactions, binary star behavior, etc.), and is likely far less predictable than what has been previously thought.
Therefore, finding hidden transient chaotic attractors represents a new and interesting challenge.

In this paper, we investigate numerically hidden transient chaotic attractors in the RF system.

The rest of the paper is organized as follows: Section 2 presents briefly the RF system and the stability of equilibria, and Section 3 deals with hidden transient chaotic attractors of the RF system. Conclusion is drawn in the last section.

\section{The RF system}

The RF system considered here is a chaotic system modeled by the following system of ODEs \cite{dan}
\begin{equation}
\label{rf}
\begin{array}{l}
\dot{x}_{1}=x_{2}\left( x_{3}-1+x_{1}^{2}\right) +ax_{1}, \\
\dot{x}_{2}=x_{1}\left( 3x_{3}+1-x_{1}^{2}\right) +ax_{2}, \\
\dot{x}_{3}=-2x_{3}\left( b+x_{1}x_{2}\right),
\end{array}%
\end{equation}

\noindent where $a>0$ and $b$ is the bifurcation parameter. The system, revealed numerically in \cite{dan}, presents unusual and extremely rich dynamics, including multistability (coexistence of multiple attractors for a given set of parameters), which represents an important ingredient for potential existence of hidden attractor.

Due to the complexity of the ODEs (third-order nonlinearities), a complete mathematical analysis such as stability of equilibria, existence of invariant sets, existence and convergence of heteroclinic or homoclinic orbits, has to be done numerically (mostly investigated in \cite{dan}).

The equilibria are $X_0^*(0,0,0)$ and
\begin{equation*}
\begin{array}{l}
X_{1,2}^{\ast }\left( \mp x^*_{1,2},\pm y_{1,2}^*,z_{1,2}^*\right),
X_{3,4}^{\ast }\left( \mp x_{3,4}^*,\pm y_{3,4}^*,z_{3,4}^*\right) ,
\end{array}%
\end{equation*}

\noindent where
\[
\begin{aligned}
x^*_{1,2}&= \pm\sqrt{\dfrac{bR_1+2b}{4b-3a}},\\  y^*_{1,2}&=\pm\sqrt{b\dfrac{4b-3a}{R_1+2}}, \\ z^*_{1,2}&=\dfrac{aR_1+R_2}{\left(4b-3a\right) R_1+8b-6a},
\end{aligned}
\]

\noindent and
\[
\begin{aligned}
x_{3,4}^*&=\pm\sqrt{\dfrac{bR_1-2b}{3a-4b}},\\y_{3,4}^*&=\pm\sqrt{b\dfrac{4b-3a}{2-R_1}},\\z_{3,4}^*&=\dfrac{aR_1-R_2}{\left(
4b-3a\right) R_1-8b+6a},
\end{aligned}
\]

\noindent with $R_1=\sqrt{3a^{2}-4ab+4}$ and $R_2=4ab^{2}-7a^{2}b+3a^{3}+2a$.

We consider in this paper the case with $a=0.1$ and $b=0.279$. Thus, the equilibria $X_{1,2,3,4}^*$ are
\[
\begin{aligned}
X_{1,2}^*&=(\mp1.1600,\pm0.2479, 0.1223),\\ \quad X_{3,4}^*&=(\mp0.0850,\pm3.3827, 0.9953).
\end{aligned}
\]

\noindent The system exhibits the symmetry $E$
\begin{equation}\label{sim}
E(x_1,x_2,x_3)\rightarrow (-x_1,-x_2,x_3).
\end{equation}

\noindent Under the transformation $E$, each trajectory has its symmetrical ("twin") trajectory with respect to the $x_3$-axis. Therefore, one can consider the stability of $X_0^*$, $X_1^*$ and $X_3^*$ only.

In order to find the hidden transient attractors, we need to determine the stability of all equilibria. The Jacobian is
\[
   J=
  \begin{bmatrix}
  2x_1x_2+a& x_1^2+x_3-1& x_2\\-3x_1^2+3x_3+1& a& 3x_1\\-2x_2x_3& -2x_1x_3& -2(x_1x_2+b)
\end{bmatrix}.
\]

\noindent All equilibria are hyperbolic.

The eigen-spectrum at the equilibrium $X_0^*$ is $\Lambda=\{\lambda_1,\lambda_2,\lambda_3\}=\{-0.5580,0.1000-1i, 0.1000+1i\}$. Thus, $X_0^*$ is a repelling focus saddle. Its two-dimensional unstable manifold $W_{X_0^*}^u=\{x_3=0\}$ (i.e. plane $(x_1,x_2)$) and the one-dimensional stable manifold $W_{X_0^*}^*=\{x_1=x_2=0\}$ (i.e. axis $x_3$), meaning that the trajectories close to $X_0^*$ move to the saddle along the axis $x_3$, but are rejected via spiralling in the plane $(x_1,x_2)$.

The equilibrium $X_1^*$ has the eigen-spectrum $\Lambda=\{-0.2470,  -0.0555 - 1.4722i,  -0.0555 + 1.4722i\}$ and, therefore, $X_1^*$ ($X_2^*$) is a stable focus node. Therefore, all trajectories approaching $X_{1,2}^*$ are attracted by these equilibria.

For the equilibrium $X_3^*$, $\Lambda=\{0.1981, -0.2780 - 4.7739i, -0.2780 + 4.7739i\}$ and, therefore, $X_3^*$ ($X_4^*$) is an attracting focus saddle. All trajectories, arriving in close vicinities of the two-dimensional stable manifold of $X_{3,4}^*$ spanned by the pair of eigenvectors corresponding to the complex conjugate eigenvalues $\lambda_{2,3}=-0.2780 \pm 4.7739i$, are attracted via spiralling toward $X_{3,4}^*$. After that, they are rejected on the unstable direction of the eigenvector corresponding to $\lambda_1=0.1981$.

Therefore, equilibria $X_{1,2}^*$ are stable and equilibria $X_0^*$, $X_{3,4}^*$ are unstable.

\section{Hidden transient chaotic attractors}

The numerical integration of the RF system represents a real challenge to ODEs solvers. The numerical results depend drastically on the initial conditions, integration step-size, and even the numerical method. Therefore, for some values of the parameters, some available fixed-step numerical methods for ODEs, implemented in software packages, might give unexpectedly different results for the same parameter values and initial conditions. Contrarily, fixed-step-size schemes, such as the standard Runge–Kutta method (RK4), or the multi-step predictor–corrector LIL method \cite{dan3}, generally give more accurate results (see details in \cite{dan}).

\begin{figure}
\includegraphics[scale=0.63]{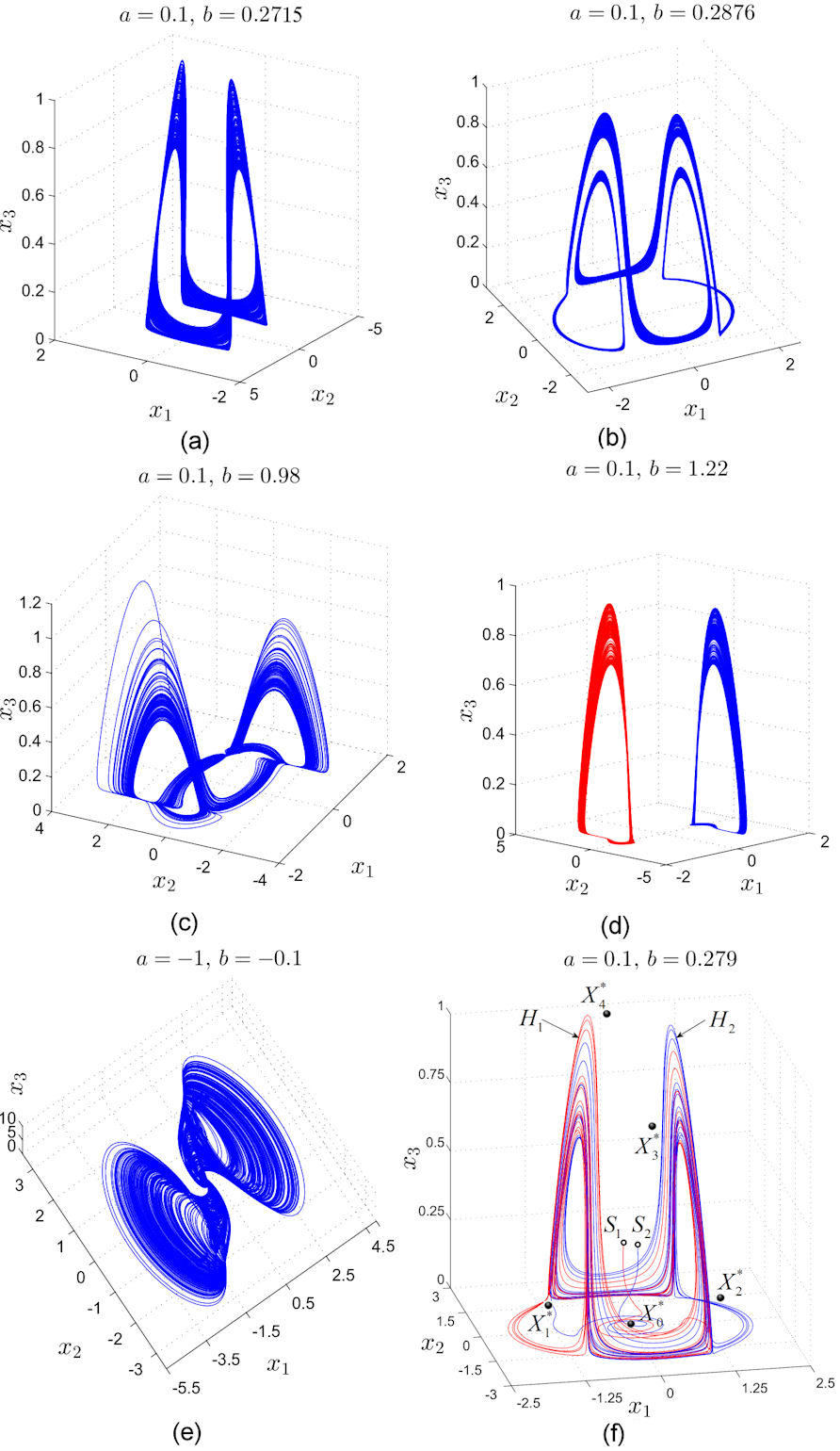}
\caption{Chaotic attractors of the RF system (\ref{rf}). (a), (b) Hidden chaotic attractors. (c) Self-excited chaotic attractor. (d) Coexisting symmetric self-excited chaotic attractors. (e) Self-exciting chaotic attractor. (f) Coexisting transient hidden chaotic attractors $H_1$ and $H_2$.}
\label{fig1}
\end{figure}

As revealed in \cite{dan}, the RF system presents extremely rich dynamics. In this paper, we are concerned with chaotic attractors only. As can be seen from Fig. \ref{fig1}, there exist different-shaped chaotic attractors and also coexisting chaotic attractors (Figs. \ref{fig1} (d),(f)). In \cite {dan2}, it is shown that, the system presents two hidden chaotic attractors corresponding to $b=0.2715$ (Fig. \ref{fig1} (a)) and $b=0.2876$ (Fig. \ref{fig1}) (b)). The other chaotic attractors (Figs. \ref{fig1} (c),(d) and (f)) are self-excited. By physical reasons, the parameters $a$ and $b$ should be positive. However, also, negative values revealed interesting dynamics (see Fig. \ref{fig1} (e), where $a=-1$ and $b=-0.1$).

We next show that the coexisting chaotic attractors ($H_1$ and $H_2$ in Fig. \ref{fig1} (f)), for $b=0.279$ and initial conditions $x_0=\pm(-0.1,0.1,0.3)$ (points $S_{1,2}$) are transient hidden chaotic attractors.

From a computational perspective and based on the connection of basins of attraction of attractors with equilibria in
the phase space, the following classification of attractors is utilized in this work.

\begin {definition}\label{deff} \cite{nick2,nick4,nick5} An attractor is called a self-excited attractor if its basin of attraction
intersects with any open neighborhood of a stationary equilibrium; otherwise, it is called hidden attractor.
\end{definition}

Due to the symmetry $E$, we only consider the attractor $H_1$.

\begin{figure}
\includegraphics[scale=0.63]{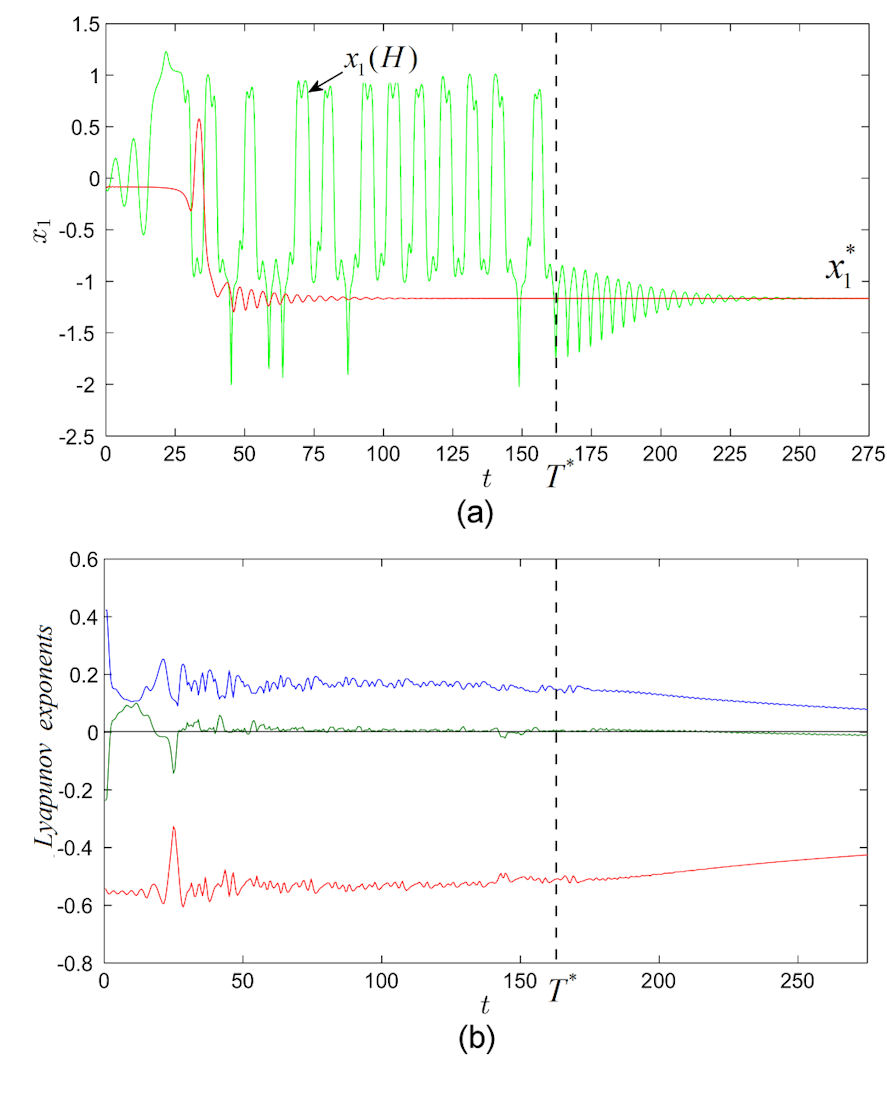}
\caption{(a) Time series of $x_1$ component of the transient hidden chaotic attractor $H_1$. (b) Lyapunov exponents of $H_1$. }
\label{fig2}
\end{figure}

The transient character of this attractor is revealed by the time series of the first component $X_1$, starting from the attraction basin of $H_1$ (Fig. \ref{fig2} (a) green plot), compared to the time series of the stable component $x_1^*$, starting from a different initial condition (red plot).

As can be seen, the lifetime of the transient is about $T^*\approx 160 s$\footnote{To obtain $T^*$ in seconds, we can use a uniform partition of the
integration time with step length corresponding to the integration
step size $h$ supposed to be measured in milliseconds.}, after which the trajectory is attracted by the stable equilibrium $X_1^*$ (component $x_1^*$ in Fig. \ref{fig2} (a)).

The chaotic behavior is underlined by the positiveness of the largest Lyapunov exponent (Fig. \ref{fig2} (b)).

\begin{figure}
\includegraphics[scale=0.8]{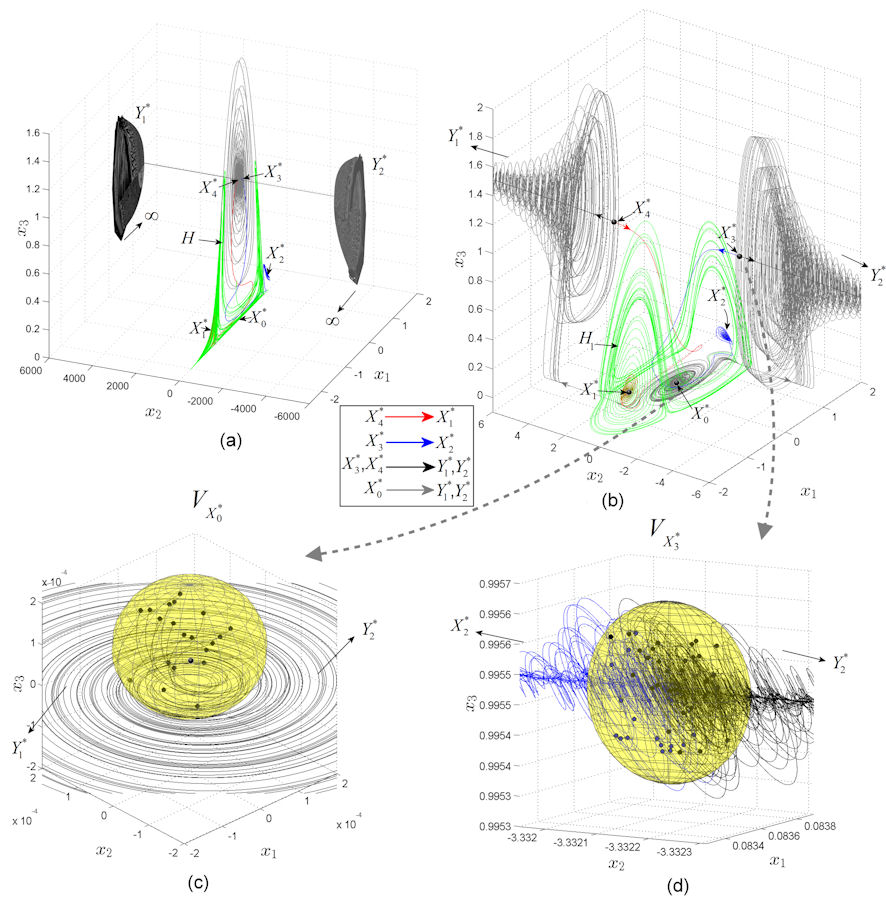}
\caption{Dynamics related to the transient hidden chaotic attractor $H_1$. (a) Hidden attractor $H_1$ and ``virtual saddles'' $Y_{1,2}^*$. (b) Zoomed image of the hidden attractor $H_1$. (c) Zoomed image of the neighborhood $V_{X_0^*}$ of the equilibrium $X_0^*$. (d) Zoomed image of the neighborhood $V_{X_{3^*}}$ of the equilibrium $X_3^*$.}
\label{fig3}
\end{figure}

In order to show that $H_1$ is a hidden transient chaotic attractor, first we have to show that all trajectories starting from vicinities of unstable equilibria $X_{3,4}^*$ tend either to stable equilibria $X_{1,2}^*$, or to infinity. In Fig. \ref{fig3} (a), the related dynamics are presented, including $H_1$ and the trajectories starting from neighborhoods of unstable equilibria.

As can be seen, unstable equilibria $X_{3,4}^*$ reject a part of the trajectories to a new kind of set, $Y_{1,2}^*$, called ``virtual saddles'' in \cite{dan}. Only few numerical methods can reveal these ``virtual saddles''. We believe that these sets are not only ``virtual'', but they really exist, and represent a characteristic of this system.

Note that the equilibria $X_{3,4}^*$, which seem to generate these ``attractors'', have their own saddles sets. Unstable directions of $X_{3,4}^*$ become stable directions in $Y_{1,2}^*$. Also, note that the distance between $Y_{1,2}^*$ and $X_{3,4}^*$ along the $x_2$ axis is quite large, in the order of $10^3$ depending on the integration step size \cite{dan,dan2}.

In order to observe more details of the shape of the transient hidden chaotic attractor $H_1$, consider the enlarge view in Fig. \ref{fig3} (b). Here, beside $H_1$ (green plot), one can see trajectories starting from unstable equilibria (for clarity only 50 trajectories are plotted).

Trajectories starting from a $\delta$-neighborhood $V_{X_0^*}$ of unstable equilibrium $X_0^*$ (with $\delta$ in the order of $10^{-4}$) tend to infinity via $Y_1^*$ or $Y_2^*$ (see grey plots in the zoomed image of $V_{X_0^*}$ in Fig. \ref{fig3} (c)). Because the plane $(x_1, x_2)$ is the unstable manifold of $X_0^*$, trajectories starting from $V_{X_0^*}$, are first attracted to this plane following the stable direction $x_3$, after which they are rejected via spiralling due the positive components of the real parts of the eigenvalues.

All trajectories starting from $\delta$-neighborhoods $V_{X_{3,4}^*}$ of $X_{3,4}^*$ (with $\delta$ in the order of $10^{-4}$) tend either to stable equilibria $X_{1,2}^*$ (red and blue plots respectively), or to infinity (black plot), via virtual saddles $Y_{1,2}^*$ (see also the zoomed vicinity $V_{X_{3}^*}$ in Fig. \ref{fig3} (d)). One can see that all trajectories exit $V_{X_{3}^*}$ via spiralling, due to the positive real parts of the complex eigenvalues $\lambda_{2,3}$ of $X_3^*$.

Summarizing, Definition \ref{deff} applies and one can conclude that $H_1$($H_2$) is a hidden transient chaotic attractor.

\begin{figure}
\includegraphics[scale=0.4]{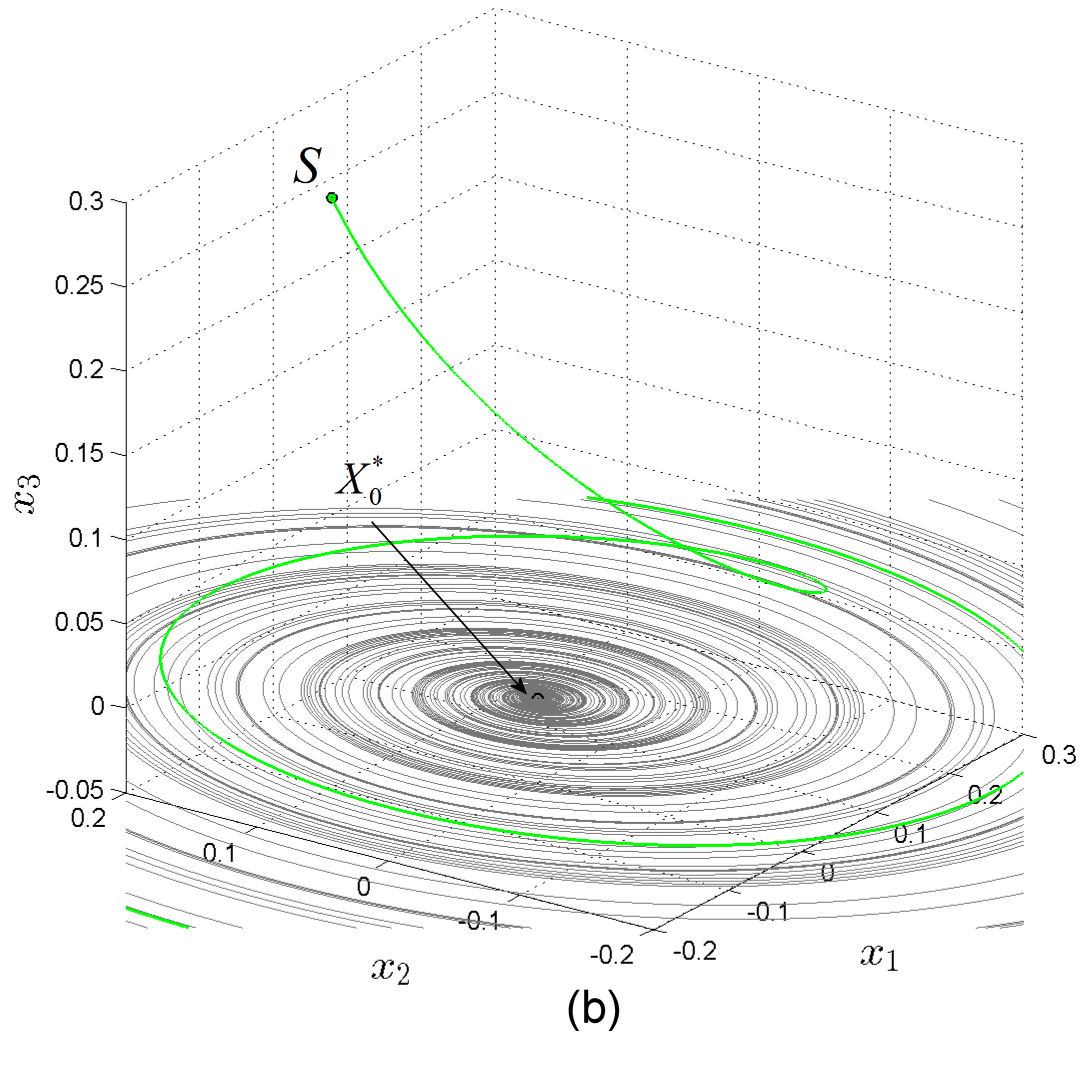}
\caption{Initial point $S$ of the transient hidden chaotic attractor.}
\label{fig4}
\end{figure}

The second task is the localization of the hidden attractor. In \cite{nick2,nick3}, an analytical-numerical localization algorithm is presented. However, trial-and-error methods could be used too. Thus, for finding $H_{1,2}$, one can utilize RK4 (or LIL algorithm) with step size $0.0001$ and initial conditions $S_{1,2}$ $x_0=\pm (-0.1,0.1,0.3)$ (Fig. \ref{fig4} reveals the position of the initial point $S_1$ situated outside the unstable neighborhoods of equilibria $X_0^*$ and $X_{3,4}^*$). Moreover, the dimension of the attraction basin of $H_1$ ($H_2$) seems to be significantly smaller than the basins of the self-excited attractors. Another modality to uncover hidden attractors is presented in \cite{kap}, where the localization is realized via perpetual points introduced in \cite{oo}.

\section{Conclusion}
In this paper, we have shown numerically that the RF system presents a pair of coexisting chaotic attractors, which are transient hidden chaotic attractors, with the lifetime of the chaotic transient $T^*\approx 160 sec$. The hidden character has been verified by mathematical definition. All trajectories starting from small neighborhoods of unstable equilibria $X_0$ and $X_{3,4}^*$ do not tend to $H_{1,2}$, but either are attracted by the stable equilibria $X_{1,2}^*$ or tend to infinity via a new kind of sets, the ``virtual saddles'' $Y_{1,2}^*$. Further studies on $Y_{1,2}^*$ will be the subject of future works. Also, studying the existence of chaotic saddles and boundary crisis in systems with transient hidden chaotic attractors, could be a good start to find if there are differences (e.g. the transient lifetime) between this kind of transient chaos and the transient chaos in the periodic windows in almost all of the chaotic systems.

\vspace{3mm}
\noindent \textbf{Acknowledgements} We thank the reviewers for their insightful comments and suggestions.


\clearpage

\clearpage

\clearpage

\clearpage

\end{document}